\documentclass{article}
\usepackage{ismir,amsmath,cite}
\usepackage{graphicx}
\usepackage{color}

\DeclareMathOperator*{\argmin}{arg\,min}

\usepackage[bookmarks=false]{hyperref}

\title{Onsets and Frames: Dual-Objective Piano Transcription}

\multauthor
{Curtis Hawthorne$^{\star}$$^{\dagger}$\thanks{$^{\star}$ Equal contribution.} \hspace{1cm} Erich Elsen$^{\star}$ \hspace{1cm} Jialin Song$^{\star}$$^{\ddagger}$\thanks{$^{\ddagger}$ Work done as a Google Brain intern.}} { \bfseries{Adam Roberts \hspace{.25cm} Ian Simon \hspace{.25cm} Colin Raffel \hspace{.25cm} Jesse Engel \hspace{.25cm} Sageev Oore \hspace{.25cm} Douglas Eck}\\
  Google Brain Team, Mountain View, CA, USA\\
{\small $^{\dagger}$ Please direct correspondance to: \tt fjord@google.com}}
\def\authorname{Curtis Hawthorne, Erich Elsen, Jialin Song, Adam Roberts, Ian Simon, Colin Raffel, Jesse Engel, Sageev Oore, Douglas Eck}

\begin{document}

\maketitle
\begin{abstract}
We advance the state of the art in polyphonic piano music transcription by using a deep convolutional and recurrent neural network which is trained to jointly predict onsets and frames.  Our model predicts pitch onset events and then uses those predictions to condition framewise pitch predictions. During inference, we restrict the predictions from the framewise detector by not allowing a new note to start unless the onset detector also agrees that an onset for that pitch is present in the frame. We focus on improving onsets \emph{and} offsets together instead of either in isolation as we believe this correlates better with human musical perception. Our approach results in over a 100\% relative improvement in note F1 score (with offsets) on the MAPS dataset. Furthermore, we extend the model to predict relative velocities of normalized audio which results in more natural-sounding transcriptions.
\end{abstract}
\section{Introduction}

Automatic music transcription (AMT) aims to create a symbolic music representation (e.g., MIDI) from raw audio. Converting audio recordings of music into a symbolic form makes many tasks in music information retrieval (MIR) easier to accomplish, such as searching for common chord progressions or categorizing musical motifs. Making a larger collection of symbolic music available also broadens the scope of possible computational musicology studies \cite{cuthbert2010music21}.

Piano music transcription is a task considered difficult even for humans due to its inherent polyphonic nature. Accurate note identifications are further complicated by the way note energy decays after an onset, so a transcription model needs to adapt to a note with varying amplitude and harmonics. Nonnegative matrix factorization (NMF) is an early popular method used in the task of polyphonic music transcription~\cite{smaragdis2003non}. With recent advancements in deep learning, neural networks have attracted more and more attention from the AMT community \cite{sigtia2016end, kelz2016potential}. In particular, the success of convolutional neural networks (CNN) for image classification tasks~\cite{szegedy2016rethinking} has inspired the use of CNNs for AMT because two-dimensional time-frequency representations (e.g., constant-Q transform~\cite{brown1991calculation}) are common input representations for audio. In \cite{kelz2016potential}, the authors demonstrated the potential for a single CNN-based acoustic model to accomplish polyphonic piano music transcription. \cite{sigtia2016end} considered an approach inspired by common models used in speech recognition where a CNN acoustic model and a Recurrent Neural Network (RNN) language model are combined. In this paper, we investigate improving the acoustic model by focusing on note onsets.

Note onset detection looks for only the very beginning of a note. Intuitively, the beginning of a piano note is easier to identify because the amplitude of that note is at its peak. For piano notes, the onset is also percussive and has a distinctive broadband spectrum. Once the model has determined onset events, we can condition framewise note detection tasks on this knowledge. Previously, \cite{cheng2016attack, wang2017two} demonstrated the promise of modeling onset events explicitly in both NMF and CNN frameworks. In this work, we demonstrate that a model conditioned on onsets achieves state of the art performance by a large margin for all common metrics measuring transcription quality: frame, note, and note-with-offset.

We also extend our model to predict the relative velocity of each onset. Velocity captures the speed with which a piano key was depressed and is directly related to how loud that note sounds. Including velocity information in a transcription is critical for describing the expressivity of a piano performance and results in much more natural-sounding transcriptions.

\section{Dataset and Metrics}
\label{dataset-metrics}

We use the MAPS dataset~\cite{emiya2010multipitch} which contains audio and corresponding annotations of isolated notes, chords, and complete piano pieces. Full piano pieces in the dataset consist of both pieces rendered by software synthesizers and recordings of pieces played by a Yamaha Disklavier player piano. We use the set of synthesized pieces as the training split and the set of pieces played on the Disklavier as the test split, as proposed in \cite{sigtia2016end}. When constructing these datasets, we also ensured that the same music piece was not present in more than one set. Not including the Disklavier recordings, individual notes, or chords in the training set is closer to a real-world testing environment because we often do not have access to recordings of a testing piano at training time. Testing on the Disklavier recordings is also more realistic because many of the recordings that are most interesting to transcribe are ones played on real pianos.

When processing the MAPS MIDI files for training and evaluation, we first translate ``sustain pedal" control changes into longer note durations. If a note is active when sustain goes on, that note will be extended until either sustain goes off or the same note is played again. This process gives the same note durations as the text files included with the dataset.

The metrics used to evaluate a model are frame-level and note-level metrics including precision, recall, and F1 score. We use the \emph{mir\_eval} library~\cite{raffel2014mir_eval} to calculate note-based precision, recall, and F1 scores. As is standard, we calculate two versions of note metrics: one requiring that onsets be within $\pm$50ms of ground truth but ignoring offsets and one that also requires offsets resulting in note durations within 20\% of the ground truth or within 50ms, whichever is greater. Frame-based scores are calculated using the standard metrics as defined in \cite{bay2009evaluation}. We also introduce a new note metric for velocity transcription that is further described in Section \ref{velocity-transcription}. Both frame and note scores are calculated per piece and the mean of these per-piece scores is presented as the final metric for a given collection of pieces.

Our goal is to generate piano transcriptions that contain all perceptually relevant performance information in an audio recording without prior information about the recording environment such as characterization of the instrument. We need a numerical measure that correlates with this perceptual goal. Poor quality transcriptions can still result in high frame scores due to short spurious notes and repeated notes that should be held. Note onsets are important, but a piece played with only onset information would either have to be entirely staccato or use some kind of heuristic to determine when to release notes. A high note-with-offset score will correspond to a transcription that sounds good because it captures the perceptual information from both onsets and durations. Adding a velocity requirement to this metric ensures that the dynamics of the piece are captured as well. More perceptually accurate metrics may be possible and warrant further research. In this work we focus on improving the note-with-offset score, but also achieve state of the art results for the more common frame and note scores and extend the model to transcribe velocity information as well.

\section{Model Configuration}
\label{model-configuration}

Framewise piano transcription tasks typically process frames of raw audio and produce frames of note activations. Previous framewise prediction models \cite{sigtia2016end, kelz2016potential} have treated frames as both independent and of equal importance, at least prior to being processed by a separate language model. We propose that some frames are more important than others, specifically the onset frame for any given note. Piano note energy decays starting immediately after the onset, so the onset is both the easiest frame to identify and the most perceptually significant.

We take advantage of the significance of onset frames by training a dedicated note onset detector and using the raw output of that detector as additional input for the framewise note activation detector. We also use the thresholded output of the onset detector during the inference process, similar to concurrent research described in \cite{thomepolyphonic}. An activation from the frame detector is only allowed to start a note if the onset detector agrees that an onset is present in that frame.

Our onset and frame detectors are built upon the convolution layer acoustic model architecture presented in \cite{kelz2016potential}, with some modifications. We use \emph{librosa}~\cite{brian_mcfee_2016_1022728} to compute the same input data representation of mel-scaled spectrograms with log amplitude of the input raw audio with 229 logarithmically-spaced frequency bins, a hop length of 512, an FFT window of 2048, and a sample rate of 16kHz. We present the network with the entire input sequence, which allows us to feed the output of the convolutional frontend into a recurrent neural network (described below).

The onset detector is composed of the acoustic model, followed by a bidirectional LSTM~\cite{schuster1997bidirectional} with 128 units in both the forward and backward directions, followed by a fully connected sigmoid layer with 88 outputs for representing the probability of an onset for each of the 88 piano keys.

The frame activation detector is composed of a separate acoustic model, followed by a fully connected sigmoid layer with 88 outputs. Its output is concatenated together with the output of the onset detector and followed by a bidirectional LSTM with 128 units in both the forward and backward directions. Finally, the output of that LSTM is followed by a fully connected sigmoid layer with 88 outputs. During inference, we use a threshold of 0.5 to determine whether the onset detector or frame detector is active.

\begin{figure}[ht]
\includegraphics[width=0.4\textwidth]{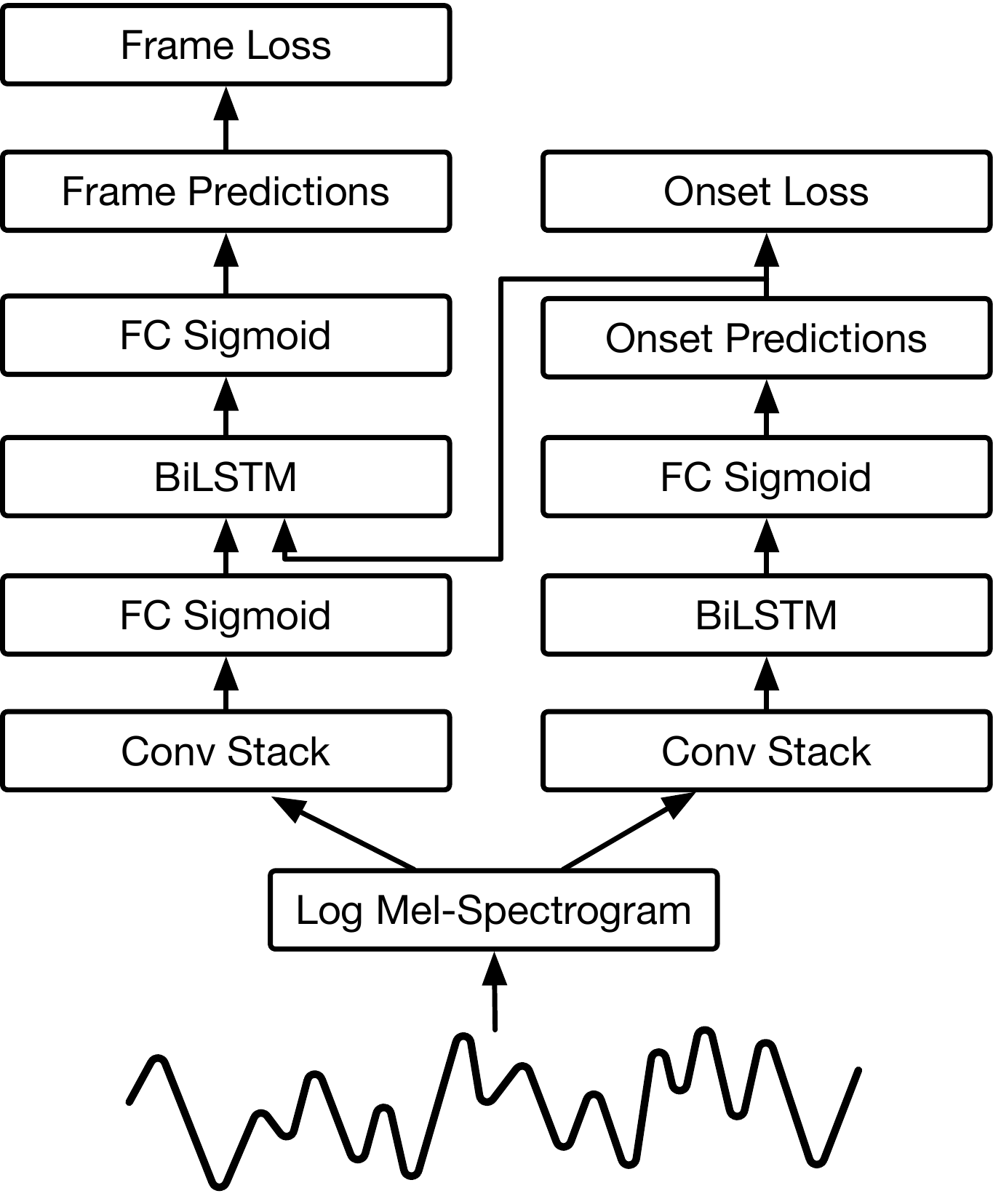}
\caption{Diagram of Network Architecture}
\end{figure}

Training RNNs over long sequences can require large amounts of memory and is generally faster with larger batch sizes. To expedite training, we split the training audio into smaller files. However, when we do this splitting we do not want to cut the audio during notes because the onset detector would miss an onset while the frame detector would still need to predict the note's presence.  We found that 20 second splits allowed us to achieve a reasonable batch size during training of at least 8, while also forcing splits in only a small number of places where notes are active.  When notes are active and we must split, we choose a zero-crossing of the audio signal.  Inference is performed on the original and un-split audio file.

Our ground truth note labels are in continuous time, but the results from audio processing are in spectrogram frames. So, we quantize our labels to calculate our training loss. When quantizing, we use the same frame size as the output of the spectrogram. However, when calculating metrics, we compare our inference results against the original, continuous time labels.

Our loss function is the sum of two cross-entropy losses: one from the onset side and one from the note side. 
\begin{equation}
L_{total} = L_{onset} + L_{frame}
\end{equation}
\begin{equation}
    L_{onset} = \sum_{p=p_{min}}^{p_{max}}\sum_{t=0}^T CE\left(\mathbf{I}_{onset}(p,t), \mathbf{P}_{onset}(p, t)\right)
\end{equation}
where $p_{min/max}$ denote the MIDI pitch range of the piano roll, $T$ is the number of frames in the example, $\mathbf{I}_{onset}(p, t)$ is an indicator function that is \texttt{1} when there is a ground truth onset at pitch $p$ and frame $t$, $\mathbf{P}_{onset}(p, t)$ is the probability output by the model at pitch $p$ and frame $t$ and $CE$ denotes cross entropy. The labels for the onset loss are created by truncating note lengths to $min(note\_length, onset\_length)$ prior to quantization. We performed a coarse hyperparameter search over $onset\_length$ (we tried 16, 32 and 48ms) and found that 32ms worked best.  In hindsight this is not surprising as it is also the length of our frames and so almost all onsets will end up spanning exactly two frames. Labeling only the frame that contains the exact beginning of the onset does not work as well because of possible mis-alignments of the audio and labels. We experimented with requiring a minimum amount of time a note had to be present in a frame before it was labeled, but found that the optimum value was to include any presence.

In addition, within the frame-based loss term $L_{frame}$, we apply a weighting to encourage accuracy at the start of the note. A note starts at frame $t_1$, completes its onset at $t_2$ and ends at frame $t_3$. Because the weight vector assigns higher weights to the early frames of notes, the model is incentivized to predict the beginnings of notes accurately, thus preserving the most important musical events of the piece. First, we define a raw frame loss as:

\begin{equation}
    L_{frame} = \sum_{p=p_{min}}^{p_{max}}\sum_{t=0}^T CE\left(\mathbf{I}_{frame}(p,t), \mathbf{P}_{frame}(p, t)\right)
\end{equation}
where $\mathbf{I}_{frame}(p, t)$ is \texttt{1} when pitch $p$ is active in the ground truth in frame $t$ and $\mathbf{P}_{frame}(p, t)$ is the probability output by the model for pitch $p$ being active at frame $t$. Then, we define the weighted frame loss as:
\begin{equation}
L_{frame}(l, p) = 
\begin{cases}
c L'_{frame}(l, p) & t_1 \leq t \leq t_2 \\
\frac{c}{t - t_2} L'_{frame} & t_2 < t \leq t_3 \\
L'_{frame}(l, p) & elsewhere
\end{cases}
\end{equation}
where $c = 5.0$ as determined with coarse hyperparameter search.

\subsection{Velocity Estimation}
\label{velocity-transcription}
We further extend the model by adding another stack to also predict velocities for each onset.  This stack is similar to the others and consists of the same layers of convolutions.  This stack does not connect to the other two.  The velocity labels are generated by dividing all the velocities by the maximum velocity present in the piece.  The smallest velocity does not go to zero, but rather to $\frac{v_{min}}{v_{max}}$.  The stack is trained with the following loss averaged across a batch:
\begin{equation}
    L_{vel} = \sum_{p=p_{min}}^{p_{max}}\sum_{t=0}^T \mathbf{I}_{onset}(p,t)(v^{p,t}_{label} - v^{p,t}_{predicted})^2
\end{equation}
At inference time the output is clipped to $[0, 1]$ and then transformed to a midi velocity by the following mapping:
\begin{equation}
    v_{midi} = 80 v_{predicted} + 10
\end{equation}

The final mapping is arbitrary, but we found this leads to pleasing audio renderings.

While various studies have considered the estimation of dynamics (note intensities or velocities) in a recording given the score \cite{szeto2005finding,ewert2011estimating,van2014predicting}, to our knowledge there has been no work in the literature considering estimation of dynamics alongside pitch and timing information. As a result, as Benetos et al.~\cite{benetos2013automatic} noted in their review paper in 2013, ``evaluating the performance of current [automatic music transcription] systems for the estimation of note dynamics has not yet been addressed."  To evaluate our velocity-aware model, we therefore propose an additional criterion for the note-level precision, recall, and F1 scores.

Evaluating velocity predictions is not straightforward because unlike pitch and timing, velocity has no absolute meaning.
For example, if two transcriptions contained identical velocities except that they were offset or scaled by a constant factor, they would be effectively equivalent despite reporting completely different velocities for every note.
To address these issues, we first re-scale all of the ground-truth velocities in a transcription to be in the range $[0, 1]$.
After notes are matched according to their pitch and onset/offset timing, we assemble pairs of the reference (ground-truth) and estimated velocities for matched notes, referred to as $v_r$ and $v_e$ respectively.
We then perform a linear regression to estimate a global scale and offset parameter such that the squared difference between pairs of reference and estimated velocities is minimized:
\begin{equation}
    m, b = \argmin_{m, b} \sum_{i = 1}^{M} \|v_r(i) - (mv_e(i) + b)\|^2
\end{equation}
where $M$ is the number of matches (i.e.\ number of entries in $v_r$ and $v_e$).
These scalar parameters are used to re-scale the entries of $v_e$ to obtain
\begin{equation}
    \hat{v}_e = \{mv_e(i) + b, i \in 1, \ldots, M\}
\end{equation}
Finally, a match $i$ is now only considered correct if, in addition to having its pitch and timing match, it also satisfies $|\hat{v}_e(i) - v_r(i)| < \tau$ for some threshold $\tau$.
We used $\tau = 0.1$ in all of our evaluations.
The precision, recall, and F1 scores are then recomputed as normal based on this newly filtered list of matches.

\section{Experiments}

We trained our onsets and frames model using TensorFlow~\cite{abadi2016tensorflow} on the training dataset described in Section~\ref{dataset-metrics} using a batch size of 8, a learning rate of .0006, and a gradient clipping L2-norm of 3.  A hyperparameter search was conducted to find the optimal learning rate. We use the Adam optimizer~\cite{kingma2014adam} and train for 50,000 steps. Training takes 5 hours on 3 P100 GPUs. The same hyperparameters were used to train all models, including those from the ablation study, except when reproducing the results of \cite{sigtia2016end} and \cite{kelz2016potential}, where hyperparameters from the respective papers were used. The source code for our model is available at \url{https://goo.gl/magenta/onsets-frames-code}.

For comparison, we reimplemented the models described in \cite{sigtia2016end, kelz2016potential} to ensure evaluation consistency. We also compared against the commercial software Melodyne version 4.1.1.011\footnote{\url{http://www.celemony.com/en/melodyne}}. We would have liked to compare against AnthemScore\footnote{\url{https://www.lunaverus.com/}} as described in \cite{troxel2016music} as well, but because it produces a MusicXML score with quantized note durations instead of a MIDI file with millisecond-scale timings, an accurate comparison was not possible.

Results from these evaluations are summarized in Table~\ref{results}. Our onsets and frames model not only produces better note-based scores (which only take into account onsets), it also produces the best frame-level scores and note-based scores that include offsets.

\begin{figure*}[htb!]
  \centerline{\includegraphics[width=.8\textwidth]{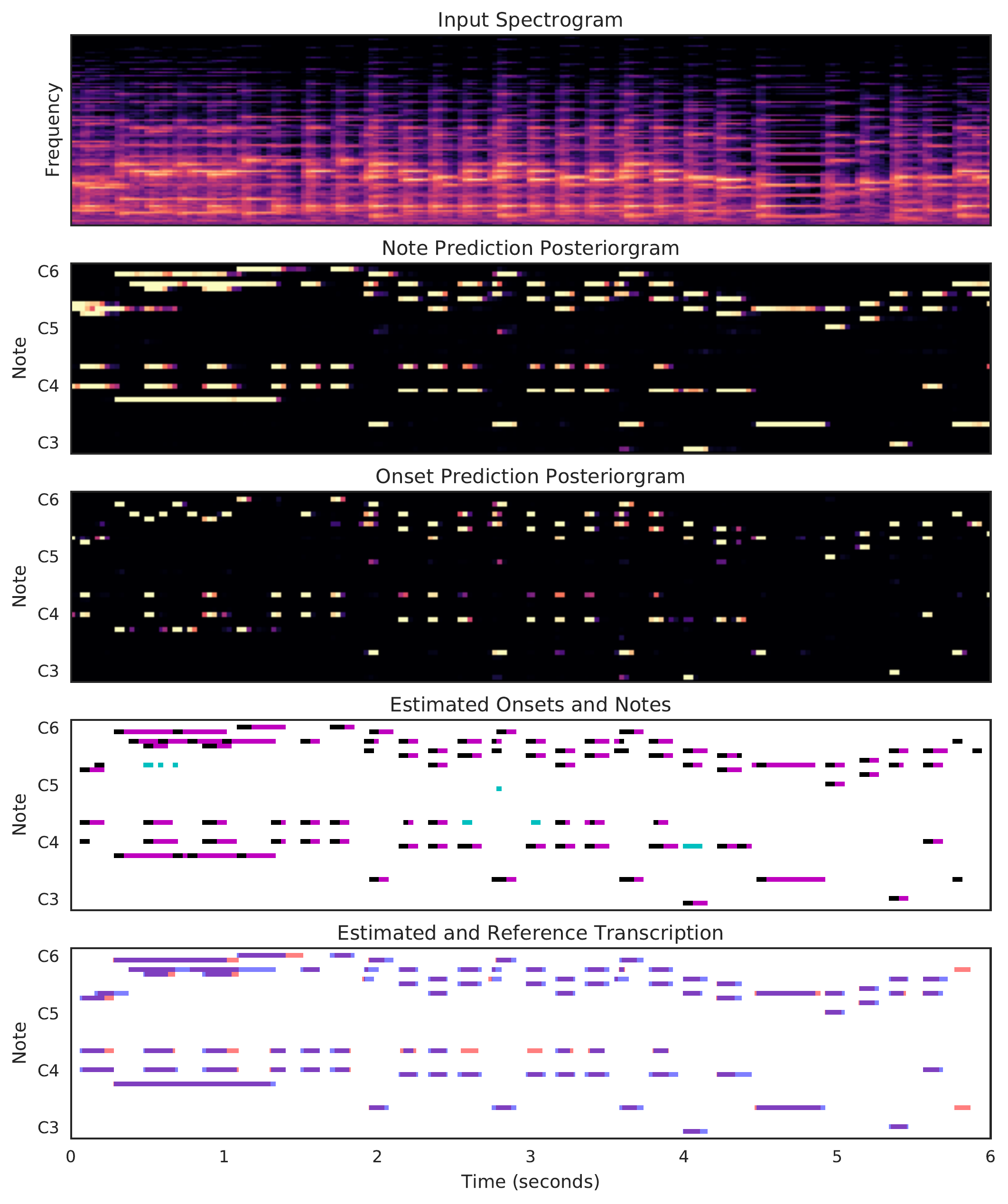}}
\caption{Inference on 6 seconds of MAPS\_MUS-mz\_331\_3\_ENSTDkCl.wav (a recording which is not in the training set).  From top to bottom, we show the log-magnitude mel-frequency spectrogram input, the framewise note probability and onset probability ``posteriorgrams'' produced by our model, the corresponding estimated onsets and notes after thresholding, and finally the resulting estimated transcription produced by our model alongside the reference transcription.  In the onset and notes plot (second from the bottom), onset predictions are shown in black.  Notes with a corresponding onset prediction are shown in magenta and notes which are filtered out because no onset was predicted for the note are shown in cyan.  In the bottom plot, the estimated transcription is shown in blue and the reference is shown in red.  Figure best viewed in color.}
\label{fig:res}
\end{figure*}

An example input spectrogram, note and onset output posteriorgrams, and inferred transcription for a recording from outside of the training set is shown in Figure~\ref{fig:res}.  The importance of restricting frame activations based on onset predictions during inference is clear: The second-to-bottom image (``Estimated Onsets and Notes'') shows the results from the frame and onset predictors. There are several examples of notes that either last for only a few frames or that reactivate briefly after being active for a while. Frame results after being restricted by the onset detector are shown in magenta. Many of the notes that were active for only a few frames did not have a corresponding onset detection and were removed, shown in cyan. Cases where a note briefly reactivated were also removed because a corresponding second onset was not detected.

Despite not optimizing for inference speed, our network performs $70\times$ faster than real time on a Tesla K40c. The MIDI files resulting from our inference experiments are available at \url{https://goo.gl/magenta/onsets-frames-examples}.

\begin{table*}[htb]
\small
\centering
\begin{tabular}{|c|c|c|c|c|c|c|c|c|c|c|c|c|}
\hline
 & \multicolumn{3}{c|}{Frame}
 & \multicolumn{3}{c|}{Note}
 & \multicolumn{3}{c|}{Note w/ offset}
 & \multicolumn{3}{c|}{Note w/ offset \& velocity}
 \\ \hline
 
 & P & R & F1 & 
 P & R & F1 & 
 P & R & F1 &
 P & R & F1
 \\ \hline
 
 Sigtia et al., 2016 \cite{sigtia2016end} &
 71.99 & \textbf{73.32} & 72.22 &
 44.97 & 49.55 & 46.58 &
 17.64 & 19.71 & 18.38 &
 \textemdash & \textemdash & \textemdash
 \\ \hline
 
 Kelz et al., 2016 \cite{kelz2016potential} &
 81.18 & 65.07 & 71.60 &
 44.27 & 61.29 & 50.94 &
 20.13 & 27.80 & 23.14 &
 \textemdash & \textemdash & \textemdash
 \\ \hline
 
 Melodyne (decay mode) &
 71.85 & 50.39 & 58.57 &
 62.08 & 48.53 & 54.02 &
 21.09 & 16.56 & 18.40 &
 10.43 & 8.15 & 9.08
 \\ \hline
 
 Onsets and Frames &
 \textbf{88.53} & 70.89 & \textbf{78.30} &
 \textbf{84.24} & \textbf{80.67} & \textbf{82.29} &
 \textbf{51.32} & \textbf{49.31} & \textbf{50.22} &
 \textbf{35.52} & \textbf{30.80} & \textbf{35.39}
 \\ \hline
 
\end{tabular}
\caption{Precision, Recall, and F1 Results on MAPS configuration 2 test dataset (ENSTDkCl and ENSTDkAm full-length .wav files). Note-based scores calculated by the \emph{mir\_eval} library, frame-based scores as defined in \cite{bay2009evaluation}. Final metric is the mean of scores calculated per piece. MIDI files used to calculate these scores are available at \protect\url{https://goo.gl/magenta/onsets-frames-examples}.}
\label{results}
\end{table*}

\section{Ablation Study}
To understand the individual importance of each piece in our model, we conducted an ablation study. We consider removing the onset detector entirely (i.e., using only the frame detector) (a), not using the onset information during inference (b), making the bi-directional RNNs uni-directional (c,d), removing the RNN from the onset detector entirely (e), pre-training the onset detector rather than jointly training it with the frame detector (f), weighting all frames equally (g), sharing the convolutional features between both detectors (h), removing the connection between the onset and frame detectors during training (i), using a Constant Q-Transform (CQT) input representation instead of mel-scaled spectrograms (j), and finally removing all the LSTMs and sharing the convolutional features (k).

These results show the importance of the onset information -- not using the onset information during inference (b) results in a significant 18\% relative decrease in the note onset score and a 31\% relative decrease in the note-with-offset score while increasing the frame score slightly.  Despite the increased frame score, the output sounds significantly worse. To our ears, the decrease in transcription quality is best reflected by the note-with-offset scores.

The model which does not have the onset detector at all (a) -- consisting of convolutions followed by a bi-directional RNN followed by a frame-wise loss -- does the worst on all metrics, although it still outperforms the baseline model from \cite{kelz2016potential}. The other ablations indicate a small impact for each component ($< 6\%$). It is encouraging that forward-only RNNs have only a small accuracy impact as they can be used for online piano transcription.

We tried many other architectures and data augmentation strategies not listed in the table, none of which resulted in any improvement. Significantly, augmenting the training audio by adding normalization, reverb, compression, noise, and synthesizing the training MIDI files with other synthesizers made no difference. We believe these results indicate a need for a much larger training dataset of real piano recordings that have fully accurate label alignments. These requirements are not satisfied by the current MAPS dataset because only 60 of its 270 recordings are from real pianos, and they are also not satisfied by MusicNet~\cite{thickstun2017learning} because its alignments are not fully accurate (e.g., there is an audible time difference between piano audio and MIDI at 1:24 in sequence 2533). Other approaches, such as seq2seq~\cite{sutskever2014sequence} may not require fully accurate alignments.

\begin{table}[htb]
\small
\centering
\begin{tabular}{|c|c|c|c|}
\hline
 & \multicolumn{3}{c|}{F1}
 \\ \hline
 
 & Frame & Note & Note \\
 & & & with offset
 \\ \hline
 
 Onset and Frames &
 \textbf{78.30} & \textbf{82.29} & \textbf{50.22}
 \\ \hline
  
 (a) Frame-only LSTM &
 76.12 & 62.71 & 27.89
 \\ \hline
 
 (b) No Onset Inference &
 \textbf{78.37} & 67.44 & 34.15
 \\ \hline
 
 (c) Onset forward LSTM &
 75.98 & 80.77 & 46.36
 \\ \hline
 
 (d) Frame forward LSTM &
 76.30 & 82.27 & 49.50
 \\ \hline
 
 (e) No Onset LSTM &
 75.90 & 80.99 & 46.14 
 \\ \hline
 
 (f) Pretrain Onsets &
 75.56 & 81.95 & 48.02
 \\ \hline
 
 (g) No Weighted Loss &
 75.54 & 80.07 & 48.55
 \\ \hline
 
 (h) Shared conv &
 76.85 & 81.64 & 43.61
 \\ \hline
 
 (i) Disconnected Detectors &
 73.91 & \textbf{82.67} & 44.83
 \\ \hline
 
 (j) CQT Input &
 73.07 & 76.38 & 41.14
 \\ \hline
 
 (k) No LSTM, shared conv &
 67.60 & 75.34 & 37.03
 \\ \hline

\end{tabular}
\caption{Ablation Study Results.}
\label{ablation-study}
\end{table}

\section{Need for more data, more rigorous evaluation}

The most common dataset for evaluation of piano transcription tasks is the MAPS dataset, in particular the ENSTDkCl and ENSTDkAm renderings of the MUS collection of pieces. This set has several desirable properties: the pieces are real music as opposed to randomly-generated sequences, the pieces are played on a real physical piano as opposed to a synthesizer, and multiple recording environments are available (``close" and ``ambient" configurations). The main drawback of this dataset is that it contains only 60 recordings. To best measure transcription quality, we believe a new and much larger dataset is needed. However, until that exists, evaluations should make full use of the data that is currently available.

Many papers, for example \cite{wang2017two, sigtia2016end, gao2017polyphonic, cogliati2017piano}, further restrict the data used in evaluation by using only the ``close" collection and/or only the first 30 seconds or less of each file. We believe this method results in an evaluation that is not representative of real-world transcription tasks.  For example, evaluating on only the ``close'' collection raises our note F1 score from 82.29 to 84.34, and evaluating on only the first 30 seconds further raises it to 86.38.  For comparison, \cite{wang2017two} achieved a note F1 score of 80.23 in this setting. The model in \cite{gao2017polyphonic} is also evaluated using 30s clips from  the ``close'' collection, but it was additionally trained on data from the test piano. This method limits the generalizability of the model but produced a note F1 score of 85.06.

In addition to the small number of the MAPS Disklavier recordings, we have also noticed several cases where the Disklavier appears to skip some notes played at low velocity. For example, at the beginning of the Beethoven Sonata No. 9, 2nd movement, several A$\flat$ notes played with MIDI velocities in the mid-20s are clearly missing from the audio (\url{https://goo.gl/magenta/onsets-frames-examples}). More analysis is needed to determine how frequently missed notes occur, but we have noticed that our model performs particularly poorly on notes with ground truth velocities below 30.

Finally, we believe that more strict metrics should be adopted by the community. As discussed in Section~\ref{dataset-metrics}, frame and note onset scores are not enough to determine whether a transcription has captured all musically relevant information from a performance. We present several audio examples at \url{https://goo.gl/magenta/onsets-frames-examples} to illustrate this point. Of the metrics currently available, we believe that the note-with-offset and velocity is the best way to compare models going forward.

Similarly, the current practice of using a 50ms tolerance for note onset correctness allows for too much timing jitter. An audio example illustrating this point is available at the above URL. We suggest future work should evaluate models with tighter timing requirements. Much work remains to be done here because as observed in \cite{bock2012polyphonic}, achieving high accuracy is increasingly difficult as timing precision is increased, in part due to the limited timing accuracy of the datasets currently available \cite{ewert2016piano}. When we trained our model at a resolution of 24ms, our scores using the existing 50ms metrics were not always as high: Frame 76.87, Note F1 82.54, Note-with-offset 49.99. Audio examples of this higher resolution model are also available at the above URL. In the examples, the higher time resolution is evident, but the model also produces more extraneous notes.

\section{Conclusion and Future Work}
We presented a jointly-trained onsets and frames model for transcribing polyphonic piano music which yields significant improvements by using onset information. This model transfers well between the disparate train and test distributions.  The current quality of our model's output is on the cusp of enabling downstream applications such as symbolic MIR and automatic music generation. To further improve the results we need to create a new dataset that is much larger and more representative of various piano recording environments and music genres for both training and evaluation. Combining an improved acoustic model with a language model is a natural next step. Another direction is to go beyond traditional spectrogram representations of audio signals.

It is very much worth listening to the examples of transcription. Consider Mozart Sonata K331, 3rd movement. Our system does a good job in terms of capturing harmony, melody, rhythm, and even dynamics. If we compare this transcription to the other systems, the difference is quite audible. We have also successfully used the model to transcribe recordings from the Musopen.org website that are completely unrelated to our training dataset. The model even works surprisingly well transcribing a harpsichord recording. Audio examples are available at \url{https://goo.gl/magenta/onsets-frames-examples}.

\section{Acknowledgements}

We would like to thank Hans Bernhard for helping with data collection and Brian McFee and Justin Salamon for consulting on the velocity metric design.

\bibliography{refs}

\end{document}